# CZTS THIN FILMS GROWN BY A SEQUENTIAL DEPOSITION OF PRECURSORS.

Daniel Cruz-Lemus[†,a], Miguel Ángel Gracia-Pinilla[†,a,b*], Mikel Hurtado[b], and Gerardo Gordillo Guzmán[c], Iver Lauermann[d]

[a] Facultad de Ciencias Físico Matemáticas, Universidad Autónoma de Nuevo León, Ciudad Universitaria, C.P 66450, San Nicolás de los Garza, México.

[b] Centro de Investigación, Innovación y Desarrollo en Ingeniería y Tecnología (CIIDIT), Universidad Autónoma de Nuevo León, C.P. 66600, Apodaca, México.

[e] Departamento de Física, Universidad Nacional de Colombia, Bogotá, Colombia.

[d] Institut für Heterogene Materialsysteme, Helmholtz-Zentrum Berlin für Materialien und Energie GmbH, Hahn-Meitner Platz 1, 14109 Berlin, Germany.

* Email Address: miguel.graciapl@uanl.edu.mx, [†] Contributed equally

ABSTRACT

A comparative study of the structural, optical and morphological properties of $Cu_2ZnSnS_4$ (CZTS) thin films prepared by two different techniques was performed. One consists of sequential evaporation of the elemental metallic precursors under a flux of sulphur supplied by evaporation from an effusion cell (physical vapor deposition-PVD) and the second one is a solution-based chemical route where thin layers of CuS, SnS and ZnS are deposited sequentially by diffusion membrane- assisted chemical bath deposition techniques; the membranes are used to optimize the kinetic growth through a moderate control of the release of the metal into CBD solution by osmosis. The present comparative study is helpful to the synthesis of kesterite nanostructured thin films.

Keywords

Kesterite, Thin Films, Raman Spectroscopy, Nanocrystals and Nanoparticles.

## 1. Introduction

Recent growth of the Cu(In,Ga)Se$_2$ and CdTe thin-film PV industry has demonstrated that thin film photovoltaic technology is becoming a viable option for large-scale power generation. However, due to the limited availability of indium and gallium, there is a concern related with the high material expenses, which restrict the capacity to lower production cost, mainly related with the desired current industrial mass production. Cu$_2$ZnSnS$_4$ has been considered one alternative absorber layer instead of Cu(In,Ga)Se$_2$ due to its earth abundant and environmental friendly elements, optimal direct band gap of 1.45eV and high absorption coefficient in the visible range[1,2]. In recent years, great efforts have been focusing on the preparation of CZTS thin films and exploration of their potential application in thin film solar cells [2-7].

Different routes have been performed to growth of thin-films, including vacuum and solution based deposition approaches. For each of these deposition approaches one key barrier towards a reliable process is the incompletely understood nature of the Cu–Zn–Sn– S phase diagram and control over the phase progression during film formation [8], which presents a challenge for preparing single phase films. A second common limitation generally encountered during process optimization involves the volatility of tin upon heating [9], which makes compositional control a challenge during film fabrication. Despite these challenges, reasonably successful film deposition and device fabrication has been demonstrated for both vacuum and solution-based deposition approaches. Vacuum-based processes include sputtering and evaporation techniques while solution based processes include electrodeposition, spray pyrolysis, and ink-based approaches [10]. Cu$_2$ZnSnSe$_4$ and Cu$_2$ZnSn(S,Se)$_4$-based solar cells using a hydrazine-based solution process have already reached an energy conversion efficiency as high as 11.1 % [11], demonstrating the effectiveness of the solution process in CZTSe-based solar cells.

In this work, we propose a novel solution based chemical route for the preparation of CZTS thin film absorbers from $Cu_2SnS_3$ and ZnS precursors and report results concerning their structural, optical and morphological properties determined from XRD, SEM, Raman spectroscopy, transmission, and AFM measurements. These are compared with those of CZTS films prepared by sequential evaporation of metallic precursors in the presence of elemental sulfur.

## 2. Experimental Setup

CZTS thin films were prepared using both, vacuum and solution based deposition techniques. The samples prepared using the vacuum based technique were grown in a PVD system equipped with two tungsten boats to evaporate copper and tin, and two effusion cells used to evaporate zinc and sulphur respectively. The temperatures of each effusion cell were controlled using PID temperature controllers and the deposition rates of both Cu and Sn were monitored with a Maxtec thickness monitor model TM-400 that uses a quartz crystal microbalance as sensor.

To grow CZTS films with a kesterite type structure, a procedure was used consisting of sequential evaporation of the metallic precursors in the presence of elemental sulphur, evaporated from an effusion cell heated around to 140°C. In order to find the optimal conditions to grow single phase CSTZ (kesterite) films, a large number of samples were deposited on glass substrates under different sequences ( Cu/Zn/Sn, Cu/Sn/Zn, Sn/Cu/Zn, Sn/Zn/Cu, Zn/Cu/Sn and Zn/Sn/Cu) varying the main deposition parameters in a wide range. XRD measurements carried out to each one of the prepared samples allowed us to find the sequence and deposition parameters which lead to the growth of $Cu_2ZnSnS_4$ thin films with kesterite structure. The study revealed that single phase CZTS films can be grown using the Cu/Sn/Zn sequence and a growth routine as that displayed in (Fig.1). Samples prepared using any other sequence or routine of growth in general showed a mixture of the CZTS phase and secondary phases like CTS.

The samples prepared using the solution based technique were grown using a procedure that consisted of sequential CBD (chemical bath deposition) steps. Deposition of thin films of $Cu_{2-x}S$, SnS and ZnS was followed by annealing at 450 °C in a nitrogen gas atmosphere. The best results were obtained by the formation of the $Cu_2ZnSnS_4$ phase in a two-step process activated during annealing. The $Cu_2SnS_3$ compound is initially formed through the chemical reaction of the $Cu_{2-x}S$ and SnS layers; subsequently, the formed $Cu_2SnS_3$ compound reacts with ZnS giving rise to the $Cu_2ZnSnS_4$ phase. The $Cu_{2-x}S$ and SnS layers were deposited by a diffusion-membrane-assisted CBD technique whereas the ZnS films were grown by the conventional CBD method [12]. Membranes are used to optimize the kinetic growth through a moderate control to release the metal into the solution by osmosis processes.

The ZnS thin films were grown in a solution containing thiourea ($CSN_2H_4$) as source of $S^{2-}$ ions, zinc acetate ($ZnC_4H_6O_4 \cdot 2H_2O$) as source of $Zn^{2+}$ ions, sodium citrate ($Na_3C_6H_5O_7 \cdot 2H_2O$) as complexing agent and ammonia ($NH_3$) for pH adjustment. The following chemical bath composition parameters led to optimal results: $[ZnC_4H_6O_4 \cdot 2H_2O]$=30mM; $[CSN_2H_4]$=400mM; $[Na_3C_6H_5O_7 \cdot 2H_2O]$=45mM. The temperature of the bath during deposition was maintained at 80°C and the solution pH around 10.0.

The synthesis of the $Cu_{2-x}S$ films was performed using a solution containing copper chloride (II) ($CuCl_2 \cdot 2H_2O$) dissolved in water as source of $Cu^{2+}$ ions, thiosulfate ($Na_2S_2O_3 \cdot 5H_2O$) as source of $S^{2-}$ ions and glucose to promote the reduction of $Cu^{2+}$ to $Cu^+$. Prior to the addition of copper chloride to the solution, this was placed on a diffusion membrane. The following chemical bath composition parameters led to optimal results: $[CuCl_2 \cdot 2H_2O]$= 37mM; $[Na_2S_2O_3 \cdot 5H_2O]$= 0,14 M; [glucose]=15mM. During the deposition the bath temperature was maintained at 80°C and the solution pH at 8.0.

The synthesis of the *$SnS_2$* films was performed using a solution containing tin chloride (II) ($SnCl_2 \cdot 2H_2O$) as source of $Sn^{2+}$ ions and thiosulfate ($Na_2S_2O_3 \cdot 5H_2O$) as source of $S^{2-}$ ions. Thiosulfate was dissolved in sodium citrate ($Na_3C_6H_5O_7 \cdot 2H_2O$) and placed on a diffusion membrane prior to the addition to the working solution;, the sodium citrate acts as complexing agent and as a pH regulator. The following chemical bath composition parameters led to optimal results: $[SnCl_2 \cdot 2H_2O]$= 133mM;

[$Na_2S_2O_3·5H_2O$]= 126mM; [$Na_3C_6H_5O_7·2H_2O$]=50mM. The temperature of the bath during was maintained at 80 °C and the solution pH around 5.0.

The transmittance measurements were done using a Varian – Cary 5000 spectrophotometer and the film thickness was determined using a Veeco Dektak 150 surface profiler. Further characterization involved X-ray diffraction with a Shimadzu-6000 diffractometer and Raman spectroscopy with a Horiba Jobin Yvon micro-Raman Spectrometer LabRamHR in backscattering configuration with a DPSS laser of 473 nm, 20 mW focused with a 50X objective. Morphological studies were also done through AFM measurements carried out with a PSI AP-0100 Scanning Probe Microscope.

### 3. Results and Discussions

The optical, structural and morphological properties of the CZTS samples grown by CBD were compared with those of reference samples containing just the $Cu_2ZnSnS_4$ phase, which were prepared by sequential evaporation of the metallic precursors in presence of elemental sulfur [12].

### 3.1 Optical properties.

The influence of the growth technique on the spectral transmittance of CuS, SnS, ZnS, $Cu_2SnS_3$ and $Cu_2SnZnS_4$ thin films was studied. In (Fig.2) typical transmission spectra of CuS, SnS, ZnS, CTS and CZTS films deposited by sequential evaporation of the metallic precursors in presence of elemental sulphur are compared with those samples prepared by sequential growth of the $Cu_xS$, SnS and ZnS precursors using a solution based chemical route.

Fig. 2 results show that both, the CZTS and CTS films exhibit low transmittance values, this behaviour is apparently caused by the presence of a high density of native defects ( vacancies, interstitial and antisite) as well as by the presence of secondary phases, which generate absorption centers within the energy gap that contribute to the photon absorption mainly in the VIS and NIR region. The CBD grown CTS films exhibit a decrease in transmittance in the IR, apparently due to the fact that this type of samples is copper rich. It is also observed that the slope of the transmittance vs $\lambda$ curves is quite small, indicating that both, the CZTS and CTS films have a poor crystallographic quality, probably associated to structural defects. It is also observed that the cut-off wavelength of the films prepared using the solution based chemical route occurs at shorter wavelengths than those prepared by evaporation. This behaviour could be attributed to the fact that the CBD prepared samples are thinner than those prepared by evaporation and additionally they are composed of several secondary phases that contribute most strongly to the photon absorption than in samples prepared by evaporation. The absorption coefficient $\alpha$ of thin films containing only the CZTS phase was determined from measurements of spectral transmittance and reflectance and using the following relationship:

$$T(\lambda) = [1 - R(\lambda)]e^{(-\alpha d)} \qquad (1)$$

Where $T(\lambda)$ is the spectral transmittance, $R(\lambda)$ the spectral reflectance, $\alpha$ the absorption coefficient and d the sample thickness.

which in its turn was used to determine the value A 600 nm thick CZTS thin film deposited by sequential evaporation under the sequence Cu /Sn /Zn was selected to get a curve of $\alpha$ vs $\lambda$ using eq. 1 and experimental values of $T(\lambda)$ and $R(\lambda)$; the values of $\alpha$ were then used to plot a curve of $(\alpha h\nu)^2$ vs $h\nu$ of $E_g$ from the intercept with the axis $h\nu$ (see inset of Fig. 2). Single phase CZTS thin films prepared by evaporation have a value of Eg = 1.42 eV and an absorption coefficient in the range $10^4$ cm$^{-1}$, indicating that they are suitable for use as the absorber layer in solar cell devices [13].

### 3.2 Structural Characterization.

Several routes have been used to prepare thin films of $Cu_2ZnSnS_4$; however, in this work we aimed to grow kesterite type CZTS thin films with tetragonal structure from $Cu_2SnS_3$ and ZnS precursors deposited sequentially in a two stage process, by both CBD and evaporation process. In a first stage the compound Cu$_2$SnS$_3$ is formed by sequential deposition of Cu$_x$S and SnS and in the second stage the compound Cu$_2$ZnSnS$_4$ is formed by deposition of ZnS on the compound formed in the first stage.

Evaporation of Cu at 550°C substrate temperature in the presence of elemental sulphur followed by evaporation of Sn at 250°C substrate temperature results in the formation of a mixture of Cu2SnS3, CuS and SnS, with the main contribution by the ternary phase . The subsequent co-evaporation of Zn and S at 550°C substrate temperature converts the binary and ternary compounds in Cu2ZnSnS4 according to the following reactions (see Fig.3).

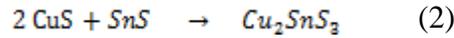  (2)

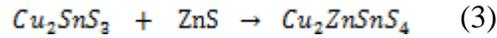  (3)

$$2\,CuS + SnS \rightarrow Cu_2SnS_3 \quad (2)$$
$$Cu_2SnS_3 + ZnS \rightarrow Cu_2ZnSnS_4 \quad (3)$$

Sequential deposition of CuxS and SnS by CBD followed by annealing at 450°C in a nitrogen atmosphere results in the formation of a mixture of Cu2SnS3, CuS and SnS. The subsequent deposition of ZnS converts the binary and ternary compounds in Cu2ZnSnS4 according to the following reactions.

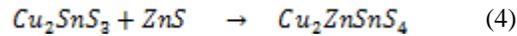  (4)

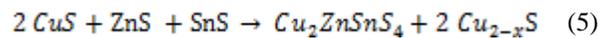  (5)

$$Cu_2SnS_3 + ZnS \rightarrow Cu_2ZnSnS_4 \quad (4)$$
$$2\,CuS + ZnS + SnS \rightarrow Cu_2ZnSnS_4 + 2\,Cu_{2-x}S \quad (5)$$

The CBD deposited CuS films grow with amorphous structure, however, analysis of the powder precipitated during the deposition by XRD gave evidence of the formation of this phase; Raman measurements confirmed this result.

Besides X-ray analysis, Raman spectroscopy measurements were done on thin films of CZTS and CTS prepared by both, sequential CBD and sequential evaporation processes. Fig. 4 shows typical room temperature Raman spectra of CTS thin films deposited by evaporation under the sequences Cu/Sn and by CBD under the sequence SnS/Cu$_x$S as well as Raman spectra of CZTS thin films deposited by evaporation under the sequences Cu/Sn/Zn and by CBD under the sequence SnS/Cu$_x$S/ZnS.

The Raman spectrum of the evaporated CZTS sample shows a single peak at 334 cm$^{-1}$ which has been attributed to the Cu$_2$ZnSnS$_4$ phase [14]. This peak arises from the A1 vibration mode of the lattice, where the group VI atom (S) vibrates while the rest of the atoms remain fixed. The Raman spectrum of the CZTS sample deposited by CBD shows a peak at 341 cm$^{-1}$ corresponding to the A1 mode of the Cu$_2$ZnSnS$_4$ phase and a second peak of lower intensity at 471 cm$^{-1}$ which has been associated with the phase Cu$_{2-x}$S [15.]. Is observed that the peak associated with the A1 mode of the sample deposited by CBD is wider than that of the evaporated sample and it occurs at a higher wavenumber. The peak broadening in the samples deposited by CBD is correlated with the increasing structural disorder due to the random distribution of S atoms in the lattice that leads to fluctuations in the masses and force constants in the neighborhood. The shift of the peak towards a higher value of the wavenumber could be associated with an enrichment of S. This behavior has been observed by other authors through study of the A1 Raman mode wavenumber dependency on the sulphur concentration in Cu$_2$ZnSn(Se$_x$S$_{1-x}$)$_4$ solid solutions [16].

The Raman spectrum of the evaporated CTS sample shows two peaks at 225 and 307 cm$^{-1}$ which have been attributed by other authors to the SnS [17] and Cu$_2$SnS$_3$ [15] phases respectively. According to the XRD measurements, this peak corresponds to the cubic Cu$_2$SnS$_3$ phase. The Raman spectrum of the CBD deposited CTS sample shows three peaks at 283, 325 and 469 cm$^{-1}$ which have been attributed to the SnS [17], Cu$_2$SnS$_3$ [15], and Cu$_{2-x}$S [15] phases respectively. The peak associated with the Cu$_{2-x}$S phase in evaporated CZTS samples is shifted with respect to that observed in CTS samples deposited by CBD, indicating that the former have a higher concentration of sulfur than the latter.

### 3.3 Morphological characterization

Fig. 5 shows AFM images corresponding to thin films of the CZTS and CTS compounds prepared by both, sequential CBD and sequential evaporation.

The AFM study revealed that the morphology of the CTS and CZTS films deposited by CBD is quite different from that of the samples prepared by sequential evaporation. The samples prepared by sequential evaporation tend to grow with grains grouped together in clusters; in CZTS films, clusters around 0.7 x1μm constituted by large grains grouped one beside the other are formed, while in CTS films the clusters are formed by stacking of laminar shape grains. CBD deposited CTS and CZTS films grow in general with a morphology characterized by small grains with sizes varying in the range of 190x 80 nm and 260x140 nm; however, in this types of samples, clusters on the order of 0.6 x0.3 μm are formed in some regions.

The SEM analysis (Fig 6) revealed the CTS and CZTS compounds present a nanometric size, also shows for both compounds the formation of micrometric aggregates, in the case of CZTS the sizes of the clusters is around of 200nm, and for the CTS is a 400nm, this sizes are similar to found by AFM study, in general correspond to typical formation of polycrystals grown by controlled of precursors deposited sequentially, we suggest that this behavior (mechanism of grown of sizes of polycrystals) can be controlled by this new route of synthesis and therefore we can regulate a aggregation and coalescence of polycrystals under solution deposition sequentially techniques.

## 4. Conclusions

CZTS thin films with tetragonal-kesterite type structure were grown from $Cu_2SnS_3$ and ZnS precursors deposited sequentially by both, vacuum and solution based deposition techniques. The vacuum based $Cu_2SnS_3$ compound was obtained by sequential evaporation of Cu and Sn in the presence of elemental sulphur and the solution based $Cu_2SnS_3$ compound was obtained using a novel procedure consisting of sequentially deposited thin films of $Cu_xS$ and SnS deposited by a diffusion-membrane-assisted CBD technique. XRD and Raman spectroscopy studies gave evidence of the formation of single phase CZTS films with kesterite structure using the vacuum based technique under optimized growth parameters; it was also found that the samples deposited using the solution based technique grow with a mixture of the $Cu_2ZnSnS_4$ and $Cu_{2-x}S$ phases.

The optical characterization performed by spectral transmittance measurements revealed that in general, the CZTS films grow with poor crystallographic quality, probably associated to structural and native defects, indicating that further studies must be done to improve the properties of the CZTS films. CZTS thin films grown in this work have a value of Eg = 1.42 eV and an absorption coefficient in the range 104 cm-1, indicating that they are suitable for use as absorber layers in solar cell devices.

The AFM and SEM study revealed that the morphology of the CTS and CZTS films deposited by CBD is quite different from that of the samples prepared by sequential co-evaporation. The samples prepared by sequential evaporation end to grow with grains grouped together in clusters, while the CBD deposited CTS and CZTS films grow in general with a morphology characterized by small grains.

ACKNOWLEDGEMENTS

This work was supported by the DIB-Universidad Nacional de Colombia and Colciencias. PAICYT-UANL grant number (11204133).